\documentclass[aps,twocolumn,showpacs]{revtex4}
\usepackage{graphicx}

\begin{document}
\title{Low-temperature hydrodynamic transport justified for dirty metal in Ioffe-Regel limit}
\author{M. V. Cheremisin}
\affiliation{A.F.Ioffe Physical-Technical Institute, St.Petersburg, Russia}
\date{\today}
\begin{abstract}
The highly disordered metals dropped into Ioffe-Regel transport regime at low-temperatures match unexpectedly the hydrodynamic criteria of shortest length scale entered the kinematic viscosity. A numerous hydrodynamic phenomena, therefore, could be of reality in this case.
\end{abstract}
\maketitle

Hydrodynamics allows to describe the transport in solid-state devices, hence, attracts a considerable attention since the
pioneering paper\cite{Gurzhi63} for 3D Poiseille flow in a wire. The recent interest to hydrodynamic description is mostly related to 2D systems\cite{Dyakonov93,Molenkamp95,Govorov04,Tomadin14,Alekseev16} including graphene\cite{Muller09,Mendoza11,Mendoza13,Torre15,Levitov16,Bandurin16,Kumar17}.
Usually, the hydrodynamic description of the electron transport is believed to occur only in ultraclean crystals at sufficiently low temperatures when
the mean-free path for carrier scattering with impurities and(or) phonons, $l_{p}$, prevails over that, $l_{ee}$, for e-e collisions. This requirement defines, in fact, a direct choice of the objects for up-to-date experimental studies.

At present, we consider the opposite case of highly disordered 3D metal falls into Ioffe-Regel\cite{Ioffe60} regime when the mean-free path $l_{p}$ becomes comparable or even less than interatomic distance, $a$. Following Refs.\cite{Wiesmann77,Gurvitch81}, starting from original scattering event the subsequent collision of electron is only possible after a finite time $a/V_{F}$ of transit between neighbors atoms. Here, $V_{F}=\hbar k_{F}/m$ is the Fermi velocity, $k_{F}$ the wave vector. This argumentation puts a basis\cite{Gurvitch81} for so-called "parallel-resistor" formula\cite{Wiesmann77}
\begin{equation}
\frac{1}{\rho}=\frac{1}{\rho(l_{p})}+\frac{1}{\rho(a)}
\label{PRF}\\
\end{equation}
for total resistivity $\rho$. Here, we use the notation $\rho(\ell)=\frac{\hbar}{e^{2}}\frac{(3\pi^{2})^{1/3}}{n^{2/3}\ell}$ with the carrier density $n$ imbedded, $\rho(l_{p})$ is the ideal resistivity followed from conventional Drude formalism. Then, $\rho(a)$ denotes the resistivity
of dirty metal in Ioffe-Regel mode under condition $na^{3}=1$\cite{Gurvitch81} equivalent to that $k_{F}=\pi/a$ proposed by Mott\cite{Mott74}. Actually, Eq.(\ref{PRF}) readily defines the total mean free path $l\sim \rho^{-1}$ as it follows
\begin{equation}
l=\frac{l_{\text {im}}l_{\text {ph}}}{l_{\text {im}}+l_{\text {ph}}}+a.
\label{Scattering_length}\\
\end{equation}
where we use the Mattissen's rule to account mean free path $l_{\text {ph(im)}}$ for phonon(impurity) scattering respectively. Usually, it is argued that Ioffe-Regel limit $l\simeq a$ becomes feasible at high temperatures($T\gg T_{\text {Debye}}$) when the phonon scattering predominates over the impurity assisted one and, therefore $l_{\text {ph}}\sim T^{-1}\rightarrow 0$. The total resistivity saturates, i.e. $\rho=\rho(a)$.

We now analyze the opposite low-temperature case($T\ll T_{\text {Debye}}$) of dumped phonon scattering when $l_{\text {ph}}\sim T^{-3}\rightarrow \infty$. According to Eq.(\ref{Scattering_length}), the total mean free path would contain a contribution caused by finite strength of static disorder, hence $l=l_{\text{im}}+a$. However, for highly disordered metal the situation is much more than dramatic. Indeed, at low temperatures the only way\cite{Gurvitch81} for electron is to be scattered on the atomic core ions separated by interatomic distance $a$. We conclude that the total mean free path must be a constant in this case, namely $l \equiv a$.

We now discuss the kinematic viscosity of Fermi gas $\eta=\frac{1}{5} V_{F}^{2}\tau_{\eta}$ introduced first in Ref.\cite{Steinberg58}. In presence of static disorder viscosity argued to depend on the second moment of the electron distribution function given by solution of Boltzmann equation, hence $\tau_{\eta}\sim\tau_{\text{im}}= a/V_{F}$. Apart from the static disorder, the e-e scattering is known to strongly influence the carrier viscosity at elevated temperatures. According to Refs.\cite{Pomeranchuk36,Quinn58,Abrikosov59} the reciprocal e-e scattering time is given by $\tau_{ee}^{-1}=\frac{(kT)^{2}}{\hbar E_{F}}$, where $E_{F}$ is the Fermi energy. Combining both contributions by means of Mattissen's rule the total length scale of viscosity transport $l_{\eta}=V_{F}\tau_{\eta}$ reads
\begin{equation}
\frac{1}{l_{\eta}}=\frac{1}{l_{ee}}+\frac{1}{a},
\label{viscosity_time}\\
\end{equation}
Here, $l_{ee}=V_{F}\tau_{ee}$ is the mean free path of e-e scattering. At $T=0$ the e-e collisions are completely damped, therefore $l_{\eta}=a$. The electron viscosity is constant $\eta \sim \frac{\hbar}{m}$. Remarkably, at finite temperature the e-e scattering invariably leads to shortening of the viscosity length, i.e. $\l_{\eta}\leq a$. One may compare viscosity length scale with that specified by Eq.(\ref{Scattering_length}). Obviously, the inequality $l_{\eta}<l$ remains always valid at finite temperature, hence provides a strong proof for hydrodynamics formalism applicability.

In conclusion, we justify the applicability of hydrodynamics for highly disordered metal in Ioffe-Regel limit at low temperatures. This theoretical study can serve as an impetus for experimental studies of the electronic hydrodynamics of dirty metals.

\end{document}